# Highly sensitive and efficient 1550 nm photodetector for room temperature operation

**RITURAJ,[1*] ZHI GANG YU,[2] R.M.E.B. KANDEGEDARA,[3] SHANHUI FAN,[4] AND SRINI KRISHNAMURTHY[2,3]**

[1]*Department of Electrical Engineering, Indian Institute of Technology Kanpur, Kanpur, U.P. 208016, India*
[2]*Sivananthan Laboratories, Bolingbrook, IL 60440, USA*
[3]*Microphysics Laboratory, University of Illinois at Chicago, Chicago, IL 60607, USA*
[4]*Department of Applied Physics, Stanford University, Stanford, CA-94305, USA*

*\*rituraj@iitk.ac.in*

**Abstract:** Photonic quantum technologies such as effective quantum communication require room temperature (RT) operating single- or few- photon sensors with high external quantum efficiency (EQE) at 1550 nm wavelength. The leading class of devices in this segment is avalanche photo detectors operating particularly in the Geiger mode. Often the requirements for RT operation and for a high EQE are in conflict, resulting in a compromised solution. We have developed a device which employs a two-dimensional (2D) semiconductor material on a co-optimized dielectric photonic crystal substrate to simultaneously decrease the dark current by three orders of magnitude at RT and maintain an EQE of >99%. The device is amenable to avalanching and form a basis for single photon detection with ultra-low dark current and high photo detection efficiency. Harnessing the high carrier mobility of 2D materials, the device has ~ps jitter time and can be integrated into a large 2D array camera.

## 1. Introduction

Detecting a low photon flux is increasingly important to the emerging fields of single-photon imaging, intensity correlation imaging, optical communication, and LIDAR in photon-starved environments [1-5]. The technologically matured CMOS detectors are readily available in the form of a large array and operate at room temperature but suffer from poor sensitivity ($\gtrsim$ 100,000 photons), low efficiency and high dark current, making them unsuitable for the aforementioned applications. On the other end of the spectrum, we have highly sensitive single-photon detectors (~ 1 photon) which are expected to be a key enabler of all large and intermediate scale photonic quantum technologies including quantum computation, communication, and sensing. It is desirable to have a room-temperature, high efficiency, fast, large array of single photon detectors. Unfortunately, the three leading technologies— superconducting nanowire single-photon detectors (SNSPDs), transition-edge sensors (TESs), and single photon avalanche diodes (SPADs)—lack in multiple of these fronts. These issues are further exacerbated — for both CMOS cameras and single-photon detectors— at the operating photon wavelength of 1550 nm. To date almost all the 1550 nm single-photon or quantum imaging demonstrations have been carried out either by raster-scanning single APD/SPAD/SNSPDs or by using small arrays with extremely slow readout, poor signal-to-noise ratio (SNR), low spatial resolution, and in limited application scenarios. Thus, there is a need to develop highly sensitive large-array cameras for detecting low fluxes of 1550 nm photons with high temporal resolution (high frame/readout rate) at room temperature. Towards this end, we propose in this work a novel detector platform utilizing the material advances in the 2D semiconductors as well as dielectric metasubstrates with tailored photonic properties. Furthermore, the proposed detector has the potential of achieving single-photon detection with all the above desired qualities.

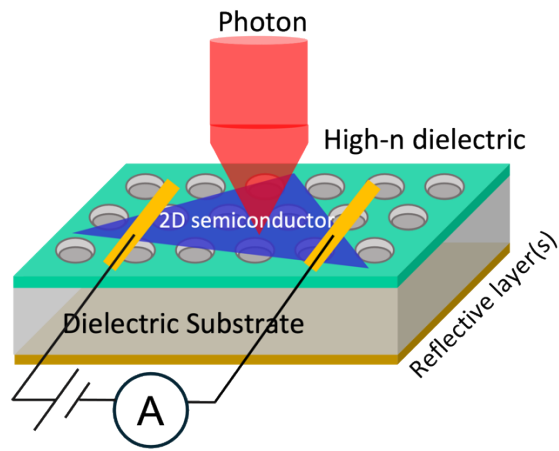

*Figure 1: Schematic design of the room-temperature high efficiency photon detector using 2D semiconductor on a suitably designed photonic crystal substrate.*

For 1550 nm-detection, the state-of-the art SPAD is InGaAs grown on InP. However, the typical dark current near breakdown is ~ 25 pA at RT [2, 22, 23]. Even with a 2.2 μm-thick and 10 μm diameter absorber yielding an EQE of 80%, the demonstrated photo detection efficiency (PDE) and dark count rate (DCR) are only ~0.3 and ~$2 \times 10^5$ respectively at RT owing to large dark current [22]. To get DCR to an acceptable range of ~$10^4$ per second, the dark current must be reduced by two orders magnitude which is often achieved by lowering the temperature to ~230 K. [22, 23, 24] The dark current can be lowered also by reducing the absorber volume, but only at the cost of lowering the EQE, and thus PDE. On the other hand, a few monolayers (MLs) of 2D materials such as $MoS_2$ and $WSe_2$, have been used as a photo detector with a dark current of $10^{-12}$ A. However, the EQE in these 2D devices is ~ 1% [3].

In this work, we consider a photodetector design consisting of a substrate, photonic crystal slab (PCS) made of transparent (to 1550 nm) material, and 2D material with a band gap of ~500 meV, as shown in Fig. 1. We show that the PCS can be designed to increase the absorption of 1550 nm photons to 100% within a monolayer while keeping the dark current in fA at RT. Further calculation of photo current-voltage (I-V) shows an SNR (defined as a ratio of photo current to dark current) >2 without carrier multiplication (avalanching) for incident light intensity of 100 nW/cm$^2$. This design can be extended to include avalanching for single-photon detection at RT. The three orders of magnitude reduction in dark current without an associated decrease in EQE could lead to high PDE and three orders of magnitude reduction in DCR when avalanching is included.

## 2. Design principle and device performance

The dark current in photo detectors arise from various recombination mechanisms such as Auger, radiative, and defect-mediated Shockley-Read-Hall mechanisms and is proportional to the absorber volume. There is a constant effort to shrink the area by requiring smaller pitch and dense arrays. To realize a high EQE the absorber thickness is usually three times the inverse of the absorption coefficient. For absorbers like InGaAs, the thickness needs to be ~ 5 μm. A reduction in thickness will linearly decrease dark current, but, unfortunately, also lower the quantum efficiency. The thickness alone cannot be reduced to get the two orders of magnitude decrease in dark current required for RT operation. On the other hand, 2D materials show considerable promise with high absorption, large mobility, and lack of surface states. For example, the photo detectors made of a few layers of 2D materials show extremely low dark currents ~pA with a EQE of about 1-3% [6-9]. Further increase in absorption while maintaining low dark current requires a clever design.

A few years ago, a PCS was designed [10] to absorb nearly 100% of 1550 nm light within one monolayer (ML) of graphene placed on it and the design was later demonstrated [11] to absorb ~90%. In these works, the PCS (square lattice of air holes in silicon) was designed to support a leaky guided resonant mode at the desired frequency which coupled to normally incident light from air. Thus, by coupling the incident light into a guided mode which travels parallel to the substrate, the authors had increased the effective path length of the light in the absorber, resulting in high absorption. Achieving 100% absorption additionally requires critical coupling condition— a delicate balance between the absorption rate in the ML and the coupling rate of the incident light into the guided resonance ($\gamma_{abs} = \gamma_c$)—to be fulfilled.

Our approach is to exploit the low dark current in 2D material and design an appropriate PCS to redirect the normally incident light to travel parallel to the 2D layer. By doing this, we obtain a longer path length and achieve a EQE of ~1 while keeping the dark current low (in pA) at RT. The graphene used in the previously published designs cannot be used for photo detection as it doesn't have a band gap and the large absorption arises from free carriers and not across a band gap. For 1550 nm operation, we require a material with a gap of 0.6 to 0.75 meV. A review of recent publications [12-14] indicated that hexagonal boron Arsenide (h-BAs) bandstructure obtained with density functional theory (DFT) and hybrid functional correction has a band gap of 0.75 eV and further calculations predicted [14] an absorption coefficient of ~$4\times10^4$ cm$^{-1}$ at 1550 nm. Additionally, h-BAs has been predicted to have high electron mobilities [15, 16] in the range of 4 to 6 $\times10^4$ cm$^2$/V.s, making it perfectly suitable for high-speed sensing (low jitter, high count rate) applications. The previous calculation of bilayer h-BAs predicted a band gap of 0.65 eV in the most stable state, but the absorption coefficients were not available.

| Stacking | AB | AB' |
|---|---|---|
| $E_g$ [eV] | 0.6 | 0.3 |
| DE [meV] | 3.2 | 0 |
| $d_{ip}$ [Å] | 1.955 | 1.955 |
| $d_{op}$ [Å] | 3.573 | 3.516 |
| $a_{1550}$[cm$^{-1}$] | $5.3\times10^4$ | $4\times10^4$ |
| $n_{1550}$ | 3.38 | 3.38 |
| $k_{1550}$ | 0.66 | 0.49 |

*Table 1: DFT values calculated for bilayer h-BAs. Band gap ($E_g$), in-plane ($d_{in}$) and intra-plane ($d_{op}$) lattice constants, absorption (a) and complex refractive index (n, k) at 1.55 µm.*

We carried out DFT calculations of bilayer h-BAs implemented through the Vienna Ab initio Simulation Package (VASP) [17]. Perdew-Burke-Ernzerhof (PBE) parametrization of the generalized gradient approximation (GGA) is used with DFT-D3 [14] to include Van der Waals forces [18]. We relax atomic coordinates and lattice constants until the intra-atomic forces are under 0.01 eV/Å. We consider two likely layer stackings in which A atom of the top layer lies on B atom of bottom layer. When the top layer is not rotated, it is called AB stack and when rotated by $60^0$, it is called AB' stack. Then we calculated the energy-dependent absorption coefficients and the complex refractive indices. The results for bilayer h-BAs are shown in Table 1. Note that AB stack is only slightly higher energy than AB' which bodes well for its growth as even higher alternate structures in h-BN and MoS$_2$ have been successfully grown [19, 20]. More importantly, this stack has a band gap of 0.60 eV and a large absorption coefficient ($\alpha$) and extinction coefficient ($k$) at 1550 nm. We used the calculated complex refractive index for further optical modeling. Although we used BAs in this work, it should be noted that the approach developed here is agnostic to the material. For example, one could as well use 3- or 4-layer phosphorene [21] in the place of BAs.

We chose Silicon as the dielectric material for designing the photonic crystal substrate due to its transparency and high refractive index at 1550 nm. The photonic crystal substrate is realized by etching periodic air holes (period *p*, radius *r*) as a square lattice in the Si substrate (thickness *t*). The geometrical parameters (*p, r, t*) are optimized to achieve a broad absorption peak centered at 1550 nm with the peak absorption approaching 100%. To protect the BAs layer from possible environmental interaction, we place BAs layer between h-BN layers as shown in Fig. 2. However, all results reported here were found to not to change with variation in the number of hexagonal boron nitride (h-BN) layers. The optimized structure (Fig. 2) with 5 to 15 hBN layers (5-15L) and its absorption spectra (Fig. 3) are shown. Figure 3a plots the absorbance for a normally incident plane wave as a function of free-space wavelength for different numbers of hBN buffer layers. The peak absorbance is close to 1 at 1.55 µm and the FWHM (full width at half maximum) linewidth is 12 nm. The broad absorption linewidth makes the device suitable for applications involving either CW (continuous wave) or pulsed photon sources. The structure is also quite robust to the variations in the thickness of the buffer layer as can be seen from the different absorption spectra in Figure 3a. Figure 3b plots the absorption spectra for different incident angles. The absorption characteristic around 1.55 µm remains practically invariant even at oblique incident angles. This broad angular absorption characteristics makes the device better amenable for realizing high resolution large array camera with small pixel size.

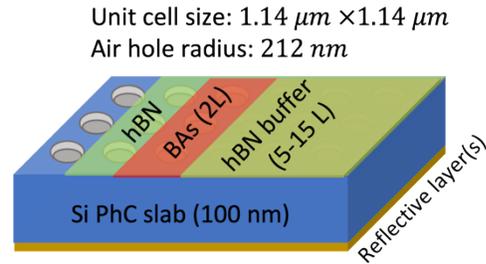

*Figure 2: Optimized 1550 nm detector structure consisting of bilayer BAs sandwiched between h-BN on a specifically designed photonic crystal substrate.*

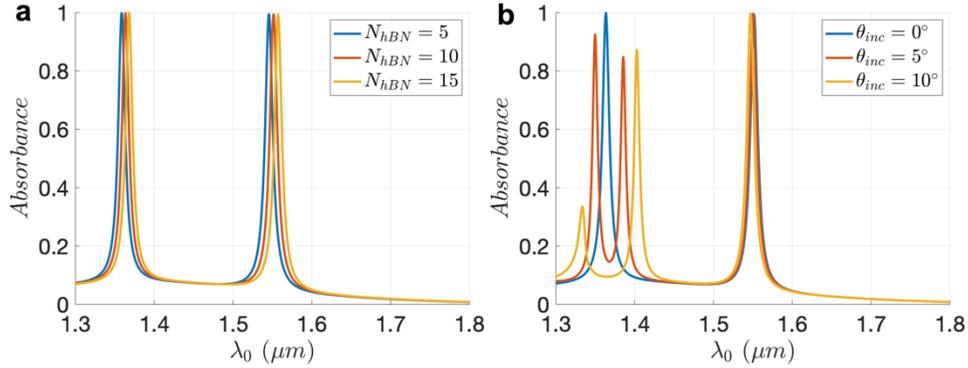

*Figure 3: Absorption spectrum a) At normal incidence (ϑ=0) for different numbers of hBN layers used to sandwich BAs absorber b) At different incident angles, as a function of free space wavelength.*

## 3. Current-voltage (I-V) calculations

After having obtained the absorption profile, we calculated the I-V characteristics of the device with and without illumination. For device simulations, we considered the 4 µm x 4 µm PiN BAs bilayer on the PCS substrate with Ohmic contacts (no Schottky barriers) on the n and p sides—device architecture is shown in Fig. 4a. The calculated absorption rate profile inside the BAs layer for a normally incident plane wave of intensity 0.1 µW/cm$^2$ at 1550 nm is shown in Fig. 4b. The dotted circle denotes the boundary of the cylindrical air hole. The radiative generation/recombination of carriers, which is non-uniform due to the underlying photonic crystal substrate, is computed based on the absorption profile in Fig. 4b. In addition, we have considered nonradiative carrier generation/recombination via the Shockley-Read-Hall (SRH) mechanism. The steady-state current-voltage (I-V) characteristics is calculated by solving drift-diffusion equations, Poisson equation and carrier continuity equations with spatially varying carrier generation-recombination, both in the absence of external illumination (dark current shown by black line in Fig. 4c) and in the presence of 0.1 µW/cm$^2$ illumination (colored line in Fig. 4c). Because of the extremely high carrier mobility expected in the BAs device, the diffusion length is much longer than the device channel length. As a result, the commonly used simple "Ohmic" boundary condition with the minority carrier density at contact being its equilibrium value is no longer valid and the proper boundary conditions should be that the gradient of minority carrier density at the contact vanishes,

$\left(\frac{dp_n}{dx}\right)_c = \left(\frac{dn_p}{dx}\right)_c = 0$, where $p_n$ ($n_p$) is the hole (electron) density in the n (p) region.

Notice that the dark current in Figure 4c, under reverse bias, is extremely small (~fA) as expected. At 0.1 V reverse bias, we predict a dark current of 5 fA in this 4 μm-square device. Since the current is directly proportional to the optical area in our fully depleted device, we would expect a current of ~25 fA if our device is changed to 10 μm-diameter device. It is instructive to compare our predicted current with the demonstrated performance of a state-of-the-art, 2.2 μm-thick and 10 μm diameter InGaAs SPAD with an EQE of 80% [22]. The measured bulk dark current at RT is 23 pA which is three orders of magnitude large than that in our proposed device—clearly indicating that reduction in the dark current arises from the reduced absorber thickness of ~1 nm BAs when compared 2.2 μm InGaAs. Consequently, we could expect the DCR to be reduced by a similar factor without any loss to high PDE. Assuming all other parameters remain unchanged, the back of the envelop calculation estimates a DCR in ~100 Hz.

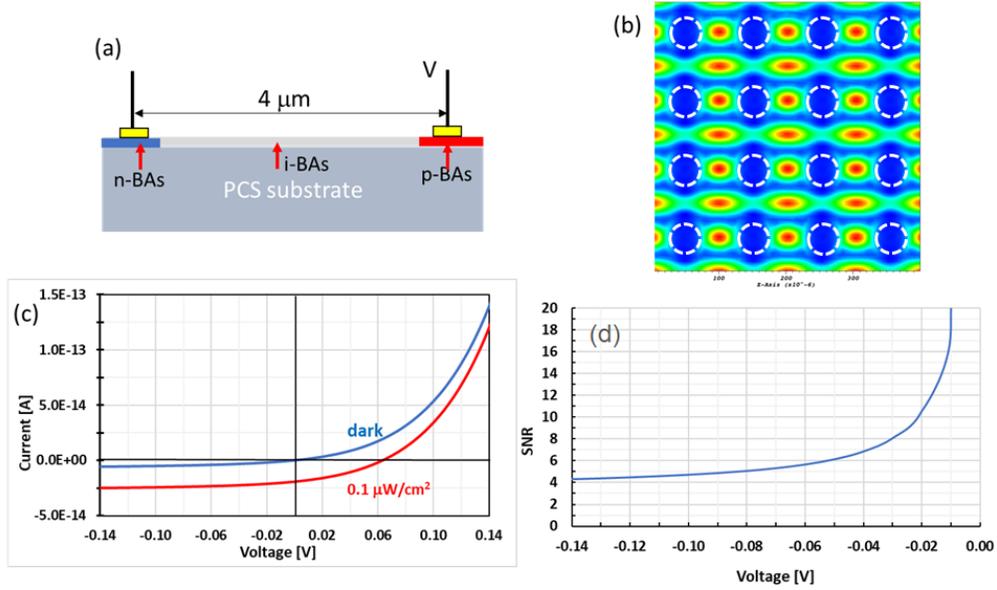

Figure 4: (a) The PiN device with bilayer BAs on a designed PCS substrate, (b) calculated absorption profile, (c) the predicted I-V curves with (red) and without (blue) 0.1 μW/cm² of 1550 nm illumination and (d) the resulting ratio of photo to dark current. The p (n) region is 0.5 μm wide with doping level of p (n) =$10^{10}$ cm$^{-2}$ and the intrinsic region is 3 μm wide with doping level of n =$10^3$ cm$^{-2}$. The SRH lifetime is set $\tau_n = \tau_p = 10^{-6}$ s. The electron and hole mobilities are $4.76 \times 10^5$ and $2.34 \times 10^5$ cm²/Vs.

When illuminated with 0.1 μW/cm² 1550 nm light, the predicted photo current is still pronounced (red line in Fig. 4c). The SNR is plotted in Fig. 4d. Notice that the SNR is very high at low bias, owing to small dark current and reaches a saturated value of ~4 at large bias. Although it is tempting to operate at low bias for a larger SNR, the intrinsic region may not be fully depleted, and the transient characteristics of the device would be very slow due to large transit time at small bias. However, 100 mV in reverse bias is large enough to fully deplete the intrinsic BAs and the SNR at that bias is ~5. Assuming a SNR of 2 is sufficient for sensing, we see that this device can sense a steady state intensity of 40 nW/cm². Since our device's area is 16 μm², in a steady state, it can sense ~53,000 photons of 1550 nm wavelength per second. This level of ultra-sensitivity at RT is possible because our design manages to achieve near 100% absorption with bilayer of absorber. Furthermore, the proposed detector structure is amenable for fabrication of a large array with small pixel pitch and high fill-factor, making it desirable for the single-photon imaging applications at 1550 nm. The underlying design principle is also suited for realizing low-photon flux cameras at mid and long wavelength IR spectrum through suitable material choices for the absorber and the PCS.

The ability to use our device for single photon detection depends on several other external factors such as RC time constant, integration time, high frequency circuitry, and measurable low current. We can make a back of the envelope estimate of various times and limitations to identify the range of the operation for single photon detection.

First, we calculate the RC time constant. The capacitance, C, of our device assuming a bilayer thickness of 1 nm, relative dielectric constant of 5 (of BAs), cross section area of 4 μm x 1 nm, and dielectric thickness of 4 μm, we get a value of $4.5 \times 10^{-20}$ F. The calculated resistance R at zero bias (Fig. 4c) is $8 \times 10^{12}$ Ω and hence RC time constant has a value of $35 \times 10^{-8}$ s or 0.35 μs.

It is instructive to calculate the transit time as well. Longer of these two times determines the rate of photon counting. The transit time is given by the ratio of length (4 μm) divided by the drift velocity which is the product of mobility (40,000 $cm^2$/V.s) and electric field (=0.1V/4μm). Substituting the values, we get 40 ps. With a larger bias, the transit time can be reduced to ~ps, leading to jitter time of similar order of magnitude. However, ~ps transit time is far smaller than the RC time. Hence RC time is used to calculate the current produced by the absorption of single photon. Since one e-h pair is created and is collected in RC time, the equivalent current is 0.5 pA. In other words, absorption of single photon gives a photo current of 0.5 pA whereas the dark current (Fig. 4c) is in fA. Equivalently, the SNR is ~ 200. However, the photon has to arrive at a rate slower than 1 per microsecond to be detected as a single photon. For faster rates, all photons impinging within a microsecond will be collectively detected. Also, external circuit has to be able to handle MHz speeds. The other concern is the ability to measure low current and the effect of the associated noise. Normally, avalanche is exploited to amplify the current. Note that our design with 4 μm of carrier travel distance will naturally include avalanching at higher bias. However, the effect of avalanching is not considered in our calculation owing to the paucity of impact ionization coefficients in 2D materials. Our future work will include time-dependent transport with and without avalanching to get the detection rates and jitter time.

In summary, we have developed a realistic design for detecting low-photon flux and possibly to a single-photon level at room temperature. The designed structure is comprised of a 2D semiconductor on a photonic crystal substrate for simultaneously achieving three order of magnitude lower dark current and high EQE. The design overcomes the inherent limiting tradeoff between photo detection efficiency and SNR in the state-of-the-art SPADs and can be readily extended to 2D materials that absorb light at other desired wavelengths. Our theoretical investigations showed that the proposed device can be both ultrasensitive and highly efficient at room temperature. The results presented here clearly establish the numerous advantages over the existing SPADs.

**Acknowledgements.** S.F. performed the work as a consultant to Sivananthan Laboratories and not part of his duties or responsibilities at Stanford University. Rituraj was supported by Sivananthan laboratories through contract No. SLI/EE/202212.

**Disclosures.** The authors declare no conflicts of interest.